\documentclass{appolb}
\usepackage{epsfig}

\newcommand{\tdm}[1]{\mbox{\boldmath $#1$}}
\def\qqd{(q\bar q)_{\rm dipole}}

\begin{document}

\title{Saturation model for 2$\gamma$ physics
\thanks{Presented by N.\ T\^\i mneanu at the X International Workshop on 
Deep Inelastic Scattering (DIS2002), Cracow, 30 April - 4 May 2002}
}
\author{N.\ T\^\i mneanu$^{a}$, J.\ Kwieci\'{n}ski$^{b}$ and  L.\ Motyka$^{a,c}$ 
\address{
$^{a}$ High Energy Physics, Uppsala University, Uppsala, Sweden   \\
$^{b}$ H.\ Niewodnicza\'{n}ski Institute of Nuclear Physics, Krak\'{o}w, Poland \\ 
$^{c}$ Institute of Physics, Jagellonian University, Krak\'{o}w, Poland\\
}}

\maketitle

\begin{abstract}
We introduce a saturation model for photon-photon interactions,
based on a QCD dipole  picture of high energy scattering. The two-dipole 
cross-section is assumed to satisfy the saturation property. This pomeron-like
contribution is supplemented
with  QPM and non-pomeron reggeon contributions. The model gives a very good 
description  of the data on the  $\gamma \gamma$ total cross-section, on the 
photon structure function $F_2^{\gamma}(x,Q^2)$ at low $x$ and on the 
$\gamma^* \gamma^*$ cross-section.
\end{abstract}

The saturation model \cite{GBW} was proven to provide a very
efficient framework to describe a variety of experimental results
on high energy scattering. With a very small number of 
free parameters, Golec-Biernat and W\"{u}sthoff (GBW) fitted 
low~$x$ data from HERA for both inclusive
and diffractive scattering \cite{GBW}. 
The central concept behind the saturation model is 
an $x$~dependent saturation scale $Q_s(x)$ at which
unitarity corrections to the linear parton evolution
in the proton become significant. 
In other words, $Q_s(x)$ is a typical scale of a hard probe 
at which a transition from a single scattering to a multiple 
scattering regime occurs.

Our idea was to extend the saturation model constructed
for $\gamma^*p$ scattering to describe also 
$\gamma^* \gamma^*$ cross sections.  
The successful extension, performed in \cite{TKM},
provided a test of the saturation model in a new
environment and confirmed the universality of the model.  
The results obtained in \cite{TKM} are also of some importance 
for two-photon physics, since the model is capable of 
describing a broad set of observables in wide kinematical
range in a simple, unified framework. 

\begin{figure}[t]
\begin{center}
\epsfig{width= 0.5\columnwidth,file=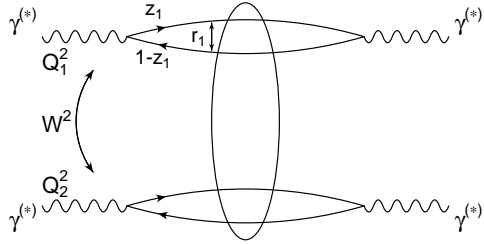}
\caption{\small\it The diagram illustrating the $\gamma^* \gamma^*$ interaction in the
dipole representation}\vspace*{-8mm}
\label{diagram}
\end{center}
\end{figure}

The saturation model for two-photon interactions
is constructed in analogy to the GBW model \cite{GBW}. 
Each of the virtual photons is decomposed into colour dipoles
$\qqd$ representing virtual components of the photon in the
transverse plane and their distribution in the photon is assumed to 
follow from the perturbative formalism (see Fig.~\ref{diagram}).

A formula for the two-photon cross-section part coming from the exchange of
{\em gluonic} degrees of freedom reads
\[
\sigma^G_{ij}(W^2,Q_1^2,Q_2^2)\; = \;
\]
\[
\sum_{a,b=1}^{N_f} \int_0^1dz_1\int d^2 {\tdm r_1}|\Psi_i^a(z_1,{\tdm r_1})|^2
\int_0^1 dz_2\int d^2 {\tdm r_2}|\Psi_j^b(z_2,{\tdm r_2})|^2
\; \sigma^{dd}_{a,b}(\bar x_{ab},r_1,r_2) ,
\]
where
$\Psi_{i}^{a} (z,{\tdm r})$ represent the photon wave functions and 
$\sigma^{dd}_{a,b}(\bar x_{ab},r_1,r_2)$ are the dipole-dipole total 
cross-sections. The indices $i,j$ label the polarisation states of the 
virtual photons, i.e.\ $T$ or $L$ and the 
different flavour content of the dipoles are specified by $a$ and $b$.
The transverse vectors ${\tdm r}_k$ denote the separation between 
$q$ and $\bar q$ in the colour dipoles and $z_k$ are the longitudinal 
momentum fractions of the quark in the photon~$k$ ($k=1,2$).

Inspired by the GBW simple choice for the dipole-proton cross-section,
we use the following parametrisation of the dipole-dipole cross-section
$\sigma_{a,b}$
\[
\sigma^{dd}_{a,b}(\bar x_{ab},r_1,r_2)\; = \;\sigma_0^{a,b}\left[
1- \exp\left(-{r_{\rm eff}^2\over  4R_0^2(\bar x_{ab})}\right)
\right],
\]
where for $\bar x_{ab}$ we take the following expression 
$\bar x_{ab} \; = \;{Q_1^2 + Q_2^2 +4m_a^2+4m_b^2\over W^2+Q_1^2+Q_2^2}$,
which allows an extension of the model down to the limit $Q_{1,2}^2=0$.
We use the same parametrisation of the saturation radius $R_0(\bar x)$
as in equation (7) in \cite{GBW}, 
and adopt the same set of parameters defining this quantity as those in
\cite{GBW}. For the saturation value $\sigma_0^{a,b}$
of the dipole-dipole cross-section we set
$\sigma_0^{a,b}\; = \;{2\over 3}\sigma_0$,
where $\sigma_0$ is the same as defined in \cite{GBW},
which  for light flavours can be justified by 
the quark counting rule in photon/proton.

For the effective separation $r_{\rm eff}(r_1,r_2)$, we consider
three scenarios, all exibiting colour transparency, i.e.\ 
$\sigma^{dd}_{a,b}(\bar x,r_1,r_2) 
\rightarrow~0$ for  $r_{1} \rightarrow 0$ or $r_{2} \rightarrow 0$:
\underline{model 1}: 
$\displaystyle r^2_{\rm  eff}\; = \;{r_1^2r_2^2 / (r_1^2+r_2^2)}$; 
\underline{model 2}:
$r^2_{\rm eff}\; = \;\min(r_1^2,r_2^2)$;\\
\underline{model 3}:
$r^2_{\rm  eff}\; = \;\min(r_1^2,r_2^2)[1+\ln(\max(r_1,r_2)/\min(r_1,r_2))]$.\\
The first two cases reduce to the original GBW model
when one of the dipoles is much larger than the other,
while option (3) is a controll case.

\begin{figure}[t]
\begin{center}
\epsfig{width= 0.6\columnwidth,file=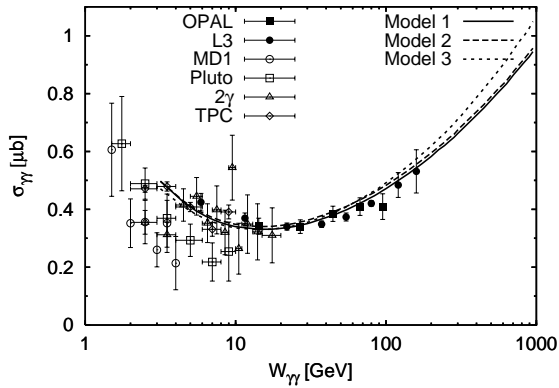} \hspace{4em} \\
\caption{\small\it The total $\gamma\gamma$ cross-section: 
data compared with all three models.}\vspace*{-8mm}
\label{real-all}
\end{center}
\end{figure}

The saturation model accounts for an exchange of {\em gluonic} degrees of
freedom, which dominate at very high energies (low~$x$). In order to get 
a complete description of $\gamma^* \gamma^*$ interactions
we should also add the non-pomeron reggeon and QPM terms \cite{JKLM},
important at lower energies.
The QPM contribution, represented by the quark box
diagrams, is well known and the cross-sections are given in \cite{BUD}.
The reggeon contribution represents a non-perturbative phenomenon
related to Regge trajectories of light mesons.
We used the parametrisation of the reggeon 
exchange cross-section in two-photon interactions from \cite{DDR}.
We have chosen the intercept in concordance with the 
value of the Regge intercept of the $f_2$ meson trajectory
$1-\eta=0.7$ \cite{PVLF0}, while other parameters
were fitted to the data on two-photon collisions. 

The formulae describing the gluonic and reggeon components are  
valid at asymptotically high energies, where
the impact of kinematical thresholds is small.
The threshold effects are approximately
accounted for by introducing a multiplicative 
correction factors, whose form is deduced
from spectator counting rules (see \cite{TKM}).
Thus, the total $\gamma^*(Q_1^2)\gamma^*(Q_2^2)$ cross-section 
reads
$\sigma_{ij} ^{\rm tot} \;=\;
\tilde\sigma_{ij} ^G  + 
\tilde\sigma ^R \delta_{iT} \delta_{jT}+
\sigma_{ij} ^{\rm QPM}$,
where $\tilde\sigma_{ij} ^G $
is the gluonic component, corresponding to dipole-dipole scattering,
with the additional
threshold correction factor.
The sub-leading reggeon $\tilde\sigma^R $
contributes only to scattering of two 
transversely polarised photons and also contains a threshold
correction; the third term $\sigma_{i,j} ^{\rm QPM}$
is the standard QPM contribution.

\begin{figure}[t]
\begin{center}
\epsfig{width= 0.49\columnwidth,file=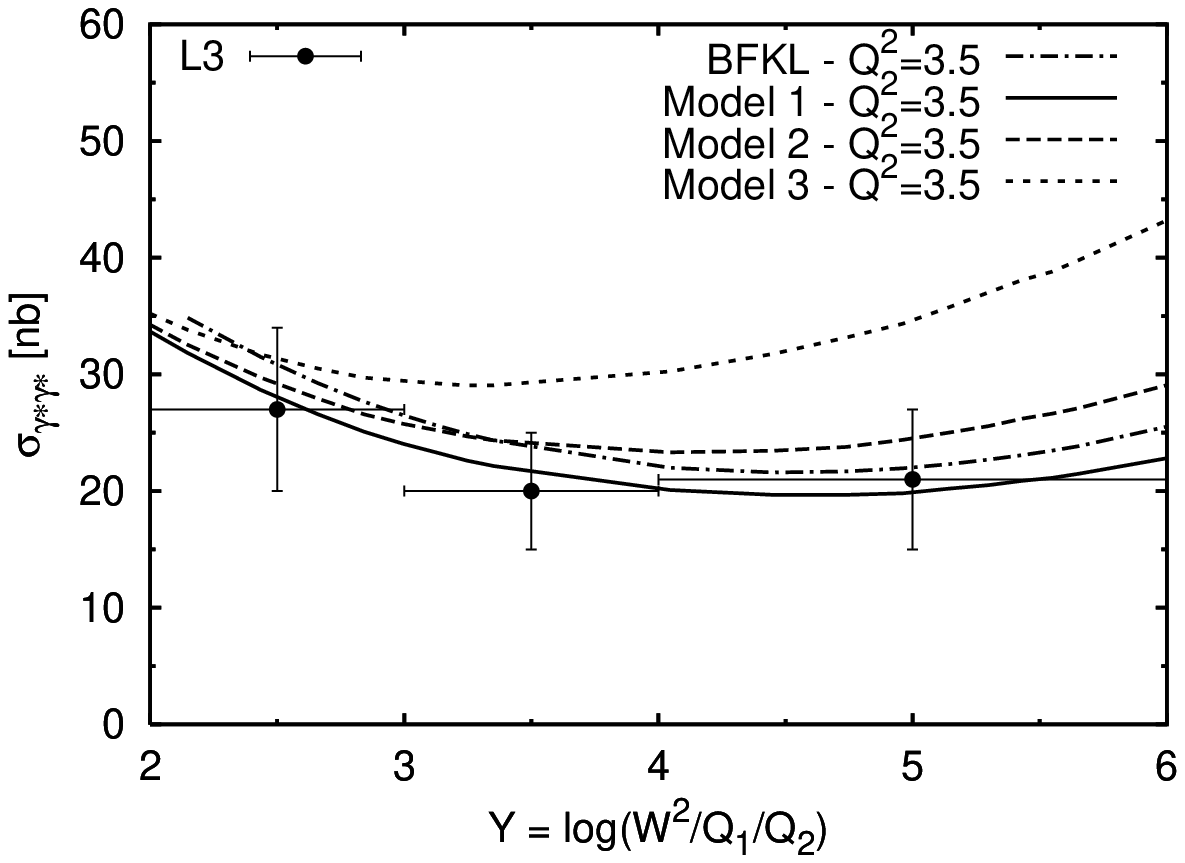}
\epsfig{width= 0.49\columnwidth,file=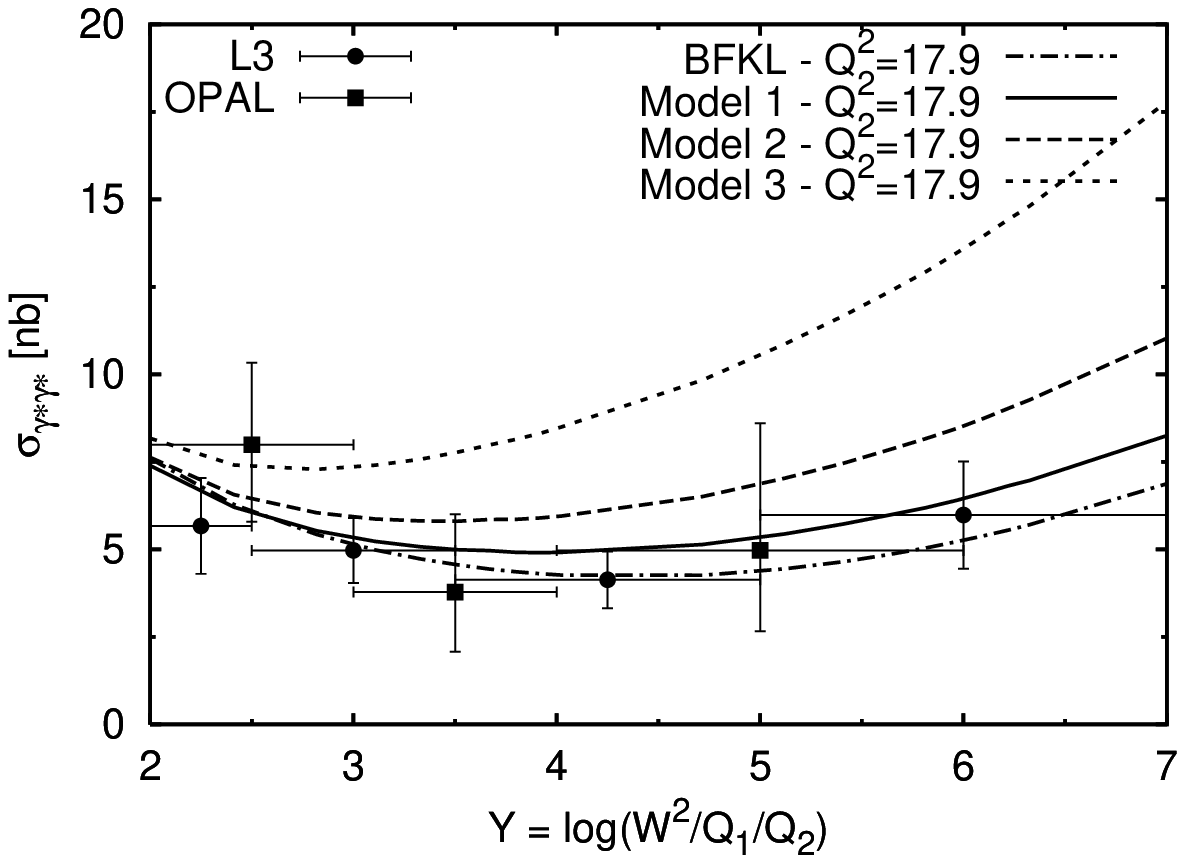}
\caption{\small\it
Total $\gamma^*\gamma^*$ cross-section for
            $Q ^2= 3.5~{\rm GeV}^2$ and 
            $Q ^2=17.9~{\rm GeV}^2$ -- comparison between LEP data 
            and the Models.
            Also shown is the result of Ref.~\cite{JKLM} based on the
            BFKL formalism with subleading corrections, supplemented by
            the QPM term, the soft pomeron and the subleading reggeon
            contributions.}\vspace*{-8mm}
\label{virt-all}
\end{center}
\end{figure}

In the comparison to the data we study three models, based on all cases
for the effective radius, as described above and we will refer to 
these models as Model 1, 2 and 3. We take without any modification the parameters
of the GBW model, however, we fit the light quark mass to the two-photon data, 
since it is not very well constrained by the GBW fit.
We find that the optimal values of the light quark ($u$, $d$ and $s$) masses
$m_q$ are 0.21, 0.23 and 0.30~GeV in Model 1,~2 and~3 correspondingly.
Also, the masses of the charm and bottom quark are tuned within the range
allowed by current measurements, to get the  optimal global description.
For the charm quark we use $m_c = 1.3$~GeV and for bottom $m_b = 4.5$~GeV.

The available data for the $\gamma\gamma$ total cross-section range
from the $\gamma\gamma$  energy $W$ equal to about 1~GeV up to
about 160~GeV,  see Fig.\ \ref{real-all}, and 
were taken for virtual photons coming from electron beams and then 
the results were extrapolated to zero 
virtualities. Some uncertainty is caused by the reconstruction
of actual $\gamma\gamma$ collision energy from the visible
hadronic energy, using an  
unfolding procedure  based on Monte Carlo programs. 
In Fig.~\ref{real-all} we show the total $\gamma\gamma$ cross-section 
from the Models, and find good agreement with data down to $W \simeq 3$~GeV
for all the Models.

\begin{figure}[t]
\begin{center}
\begin{tabular}{cc}
\epsfig{width= 0.48\columnwidth,file=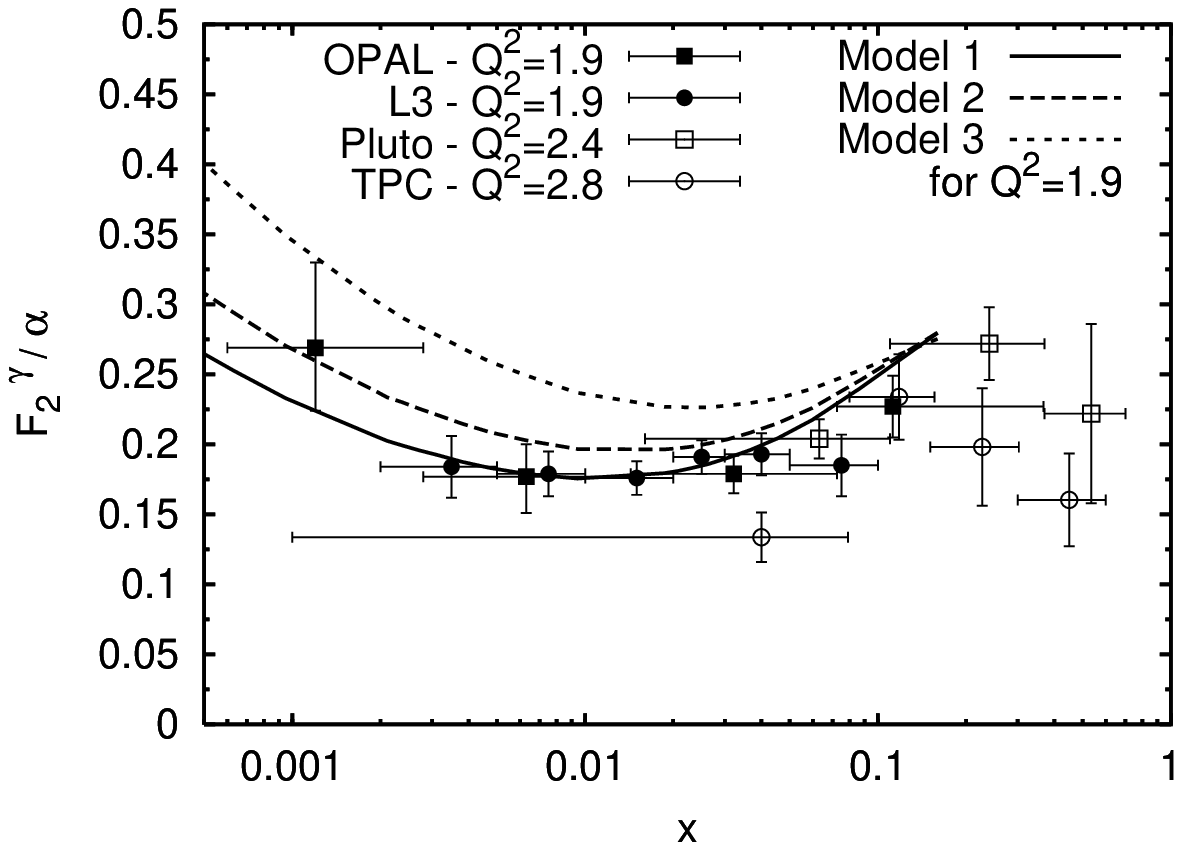} &
\epsfig{width= 0.48\columnwidth,file=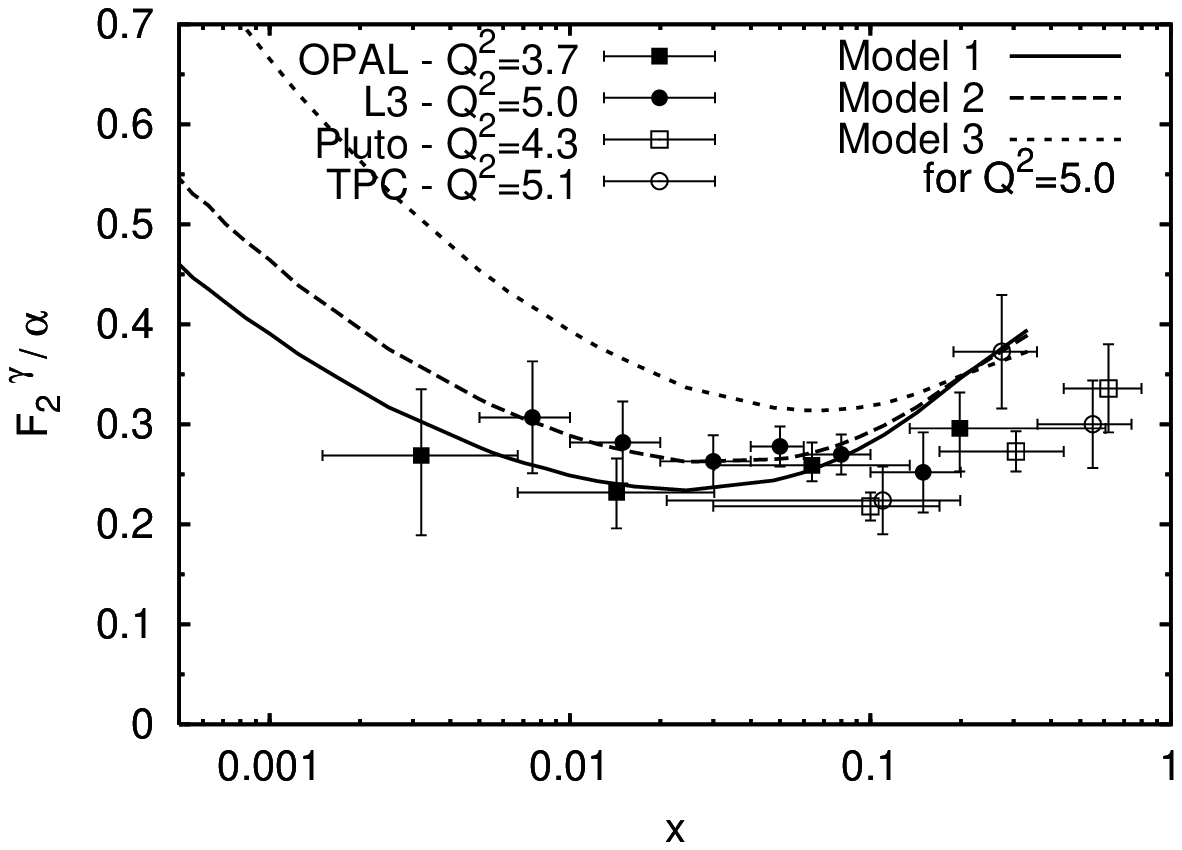}\\
\epsfig{width= 0.48\columnwidth,file=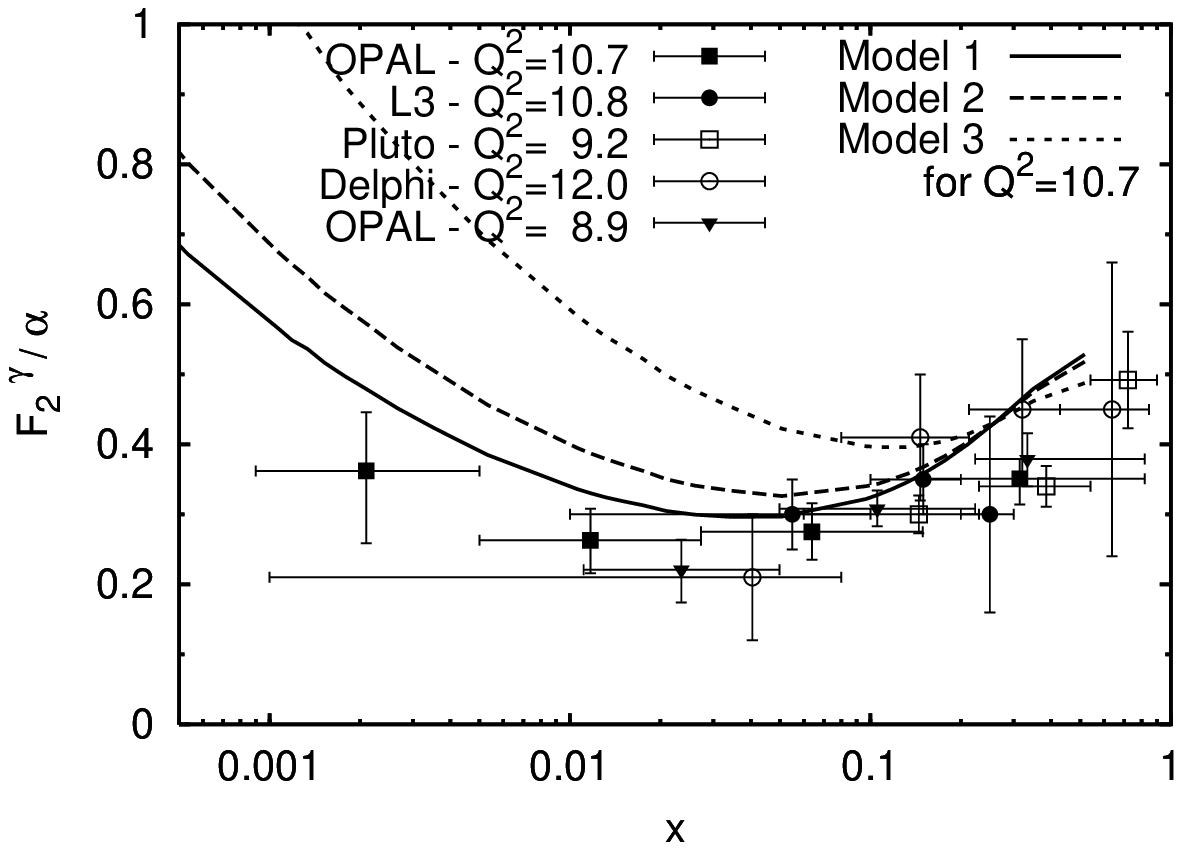} &
\epsfig{width= 0.48\columnwidth,file=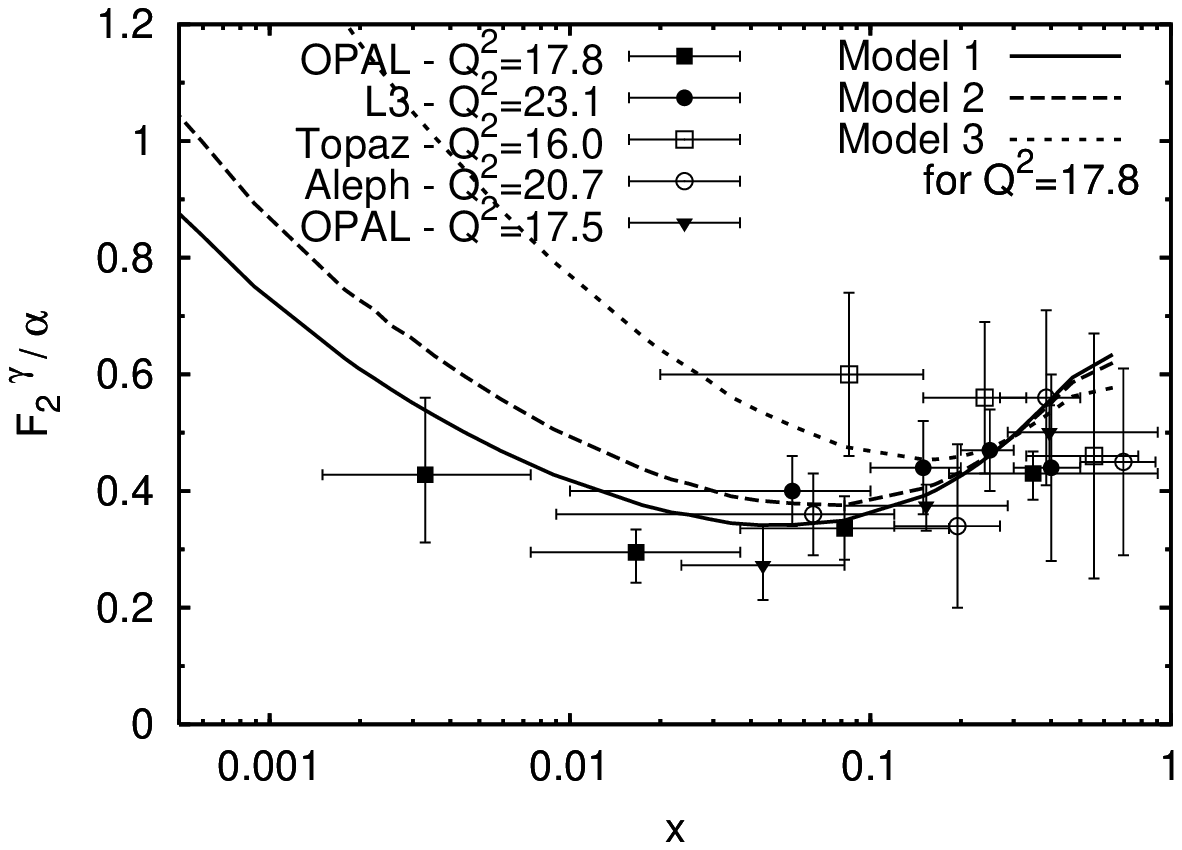} \\
\end{tabular}
\caption{\small\it The photon structure function $F_2 ^{\gamma}(x,Q^2)$:
the experimental data compared to predictions following from the
Models for various $Q^2$ values.}\vspace*{-8mm}
\label{f2-all}
\end{center}
\end{figure}

The data for the total $\gamma^*\gamma^*$ cross-section are extracted
from so-called double-tagged events: $e^+e^-$ events
in which both scattered electrons are measured.
In such events  measurement of the kinematical variables of the leptons
determines both the virtualities $Q_1^2$ and $Q_2^2$ of the colliding photons and
the collision energy $W$. 
In Fig.~\ref{virt-all} these data are compared with the curves from 
the Models. Models 1 and~2 fit the data well whereas Model~3 does not.
The virtuality of both photons are large, so the unitarity corrections, 
the light quark mass effects and the reggeon contribution are not 
important here. Moreover, the perturbative approximation for the photon 
wave function is fully justified in this case. 
Thus, in this measurement the form of the dipole-dipole
cross-section is directly probed.

The data on quasi-real photon structure are obtained 
mostly in single tagged $e^+e^-$ events, in which
a two-photon collision occurs. One of the photons has 
a large virtuality and probes the other, almost real photon.
In Fig.~\ref{f2-all}  we show  the comparison of our predictions with the 
experimental data.
Model 1, favoured by the $\gamma^*\gamma^*$ data provides the best description
of $F_2 ^{\gamma}$ as well. 

In conclusion, our extension of the saturation approach to two photon
physics provides a simple and efficient framework to calculate observables
in $\gamma\gamma$ processes and good agreement has been found with the available data.

\end{document}